\makeatletter
\newcommand{\smallsym}[2]{#1{\mathpalette\make@small@sym{#2}}}
\newcommand{\make@small@sym}[2]{\vcenter{\hbox{$\m@th\downgrade@style#1#2$}}}
\newcommand{\downgrade@style}[1]{\ifx#1\displaystyle\scriptstyle\else
    \ifx#1\textstyle\scriptstyle\else
      \scriptscriptstyle
  \fi\fi
}
\makeatother

\documentclass{edm_article}

\usepackage{hyperref}

\begin{document}

\title{Uncertainty-preserving deep knowledge tracing with state-space models} 

\numberofauthors{3}
\author{
\alignauthor
S. Thomas Christie\\
       \affaddr{Carleton College}\\
       \affaddr{Northfield, MN 55057}\\
       \email{tchristie@carleton.edu}
\alignauthor
Carson Cook\\
       \affaddr{NWEA/HMH}\\
       \affaddr{Beaverton, Oregon, 97005}\\
       \email{carsonjamescook@gmail.com}
\alignauthor
Anna N. Rafferty\\
       \affaddr{Carleton College}\\
       \affaddr{Northfield, MN 55057}\\
       \email{arafferty@carleton.edu}
}

\maketitle

\begin{abstract}
A central goal of both knowledge tracing and traditional assessment is to quantify student knowledge and skills at a given point in time. Deep knowledge tracing flexibly considers a student's response history but does not quantify epistemic uncertainty, while IRT and CDM compute measurement error but only consider responses to individual tests in isolation from a student's past responses. Elo and BKT could bridge this divide, but the simplicity of the underlying models limits information sharing across skills and imposes strong inductive biases. To overcome these limitations, we introduce Dynamic LENS, a modeling paradigm that combines the flexible uncertainty-preserving properties of variational autoencoders with the principled information integration of Bayesian state-space models. Dynamic LENS allows information from student responses to be collected across time, while treating responses from the same test as exchangeable observations generated by a shared latent state. It represents student knowledge as Gaussian distributions in high-dimensional space and combines estimates both within tests and across time using Bayesian updating. We show that Dynamic LENS has similar predictive performance to competing models, while preserving the epistemic uncertainty -- the deep learning analogue to measurement error -- that DKT models lack. This approach provides a conceptual bridge across an important divide between models designed for formative practice and summative assessment.

\end{abstract}

\keywords{knowledge tracing, assessment, information integration} \section{Introduction}

During the course of a school year, students produce a stream of behavioral evidence that reflects changes in their knowledge and skills. Students do daily homework, regular in-class formative assessments, and take multiple standardized assessments each year. A teacher could integrate this behavioral evidence to both interpret the student's progress and needs, and simultaneously recognize when more information is needed to make confident decisions.  Unfortunately, current mathematical and computational models are insufficient to handle this dual mandate: most \textit{assessment} models do not integrate information over time, and deep knowledge tracing models do not capture the epistemic uncertainty necessary to know when predictions should be trusted.  Holistically integrating student behavioral evidence across formative and summative interactions, while simultaneously tracking uncertainty and confidence, would allow past measurements to inform current measurements in a principled way. This capability could be used to shorten standardized assessments, increase the reliability of formative assessments, and reduce the burdens associated with retesting.

\begin{figure*}
    \centering
    \includegraphics[width=1\linewidth]{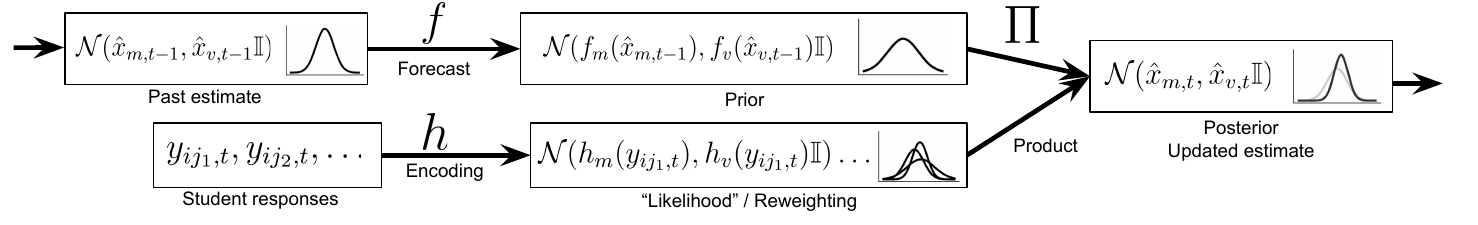}
    \caption{The dynamic LENS model. A learned forecasting function updates beliefs about the student latent state over time, and a learned encoding re-weights this distribution to incorporate new evidence.}
    \label{fig:conceptual}
    \Description{A diagram illustrating the computational flow of the Dynamic LENS model. Past estimates of student skill are forecast into the present, resulting in a normal distribution treated as a prior for the present timestep. Student responses are also encoded into normal distribution and treated as a likelihood. These are multiplied to create a normal distribution treated as a posterior belief of student skills at time t.}
\end{figure*}

One challenge for integrating such information is that different paradigms are used for designing and scoring different types of assessments. Interim and summative tests treat assessment as \textit{measurement}, using latent variable models like Item Response Theory~\cite{chen2021irt} to estimate latent parameters associated with student skills, and reporting results with corresponding measurement error. Yet, homework and other formative practice is scored very differently, for example by completion or detecting several correct responses in a row. While each venue is an opportunity for students to exhibit their knowledge and growth over time, the different scoring mechanisms and types of information obtained mean that results are often treated in isolation and are difficult to integrate.

Ideally, a computational model could combine multiple types of evidence into a coherent and detailed portrait of student learning. In recent years, Deep Knowledge Tracing (DKT) models have been introduced that promise to flexibly integrate information from observed student behavior across time~\cite{piech2015dkt, lee2019kqn, nakagawa2019gkt, pandey2019sakt, shin2021saint}. 
DKT models use machine learning techniques like recurrent neural networks and, more recently, transformer and graph architectures~\cite{piech2015dkt, nakagawa2019gkt, pandey2019sakt, shin2021saint} to predict student performance. 
DKT models can be viewed as modern variants of Bayesian Knowledge Tracing (BKT, \cite{corbett1994knowledge}) and Elo~\cite{elo}, which similarly integrate information across time to make inferences about student progress. 

While DKT models are flexible and often highly predictive, they do not model epistemic uncertainty about student knowledge states. 
Epistemic uncertainty is uncertainty that can be reduced by gathering more observations, such as the uncertainty associated with an estimate of a coin's bias that decreases as more flips are observed~\cite{hullermeier2021aleatoricepistemic}. In measurement models like IRT, epistemic uncertainty is a key concern. Computer adaptive tests (CAT) rely on measurement of epistemic uncertainty to select questions~\cite{wainer2000computerized}. 
Epistemic uncertainty also determines test length in CATs: students respond to questions until epistemic uncertainty (framed as measurement error) is reduced below a threshold value. However, models like IRT are not as flexible as DKT. IRT models treat the space of student skills as very low dimensional, do not integrate information over time, and make strong assumptions about the form of latent student distributions. 

We introduce a model called Dynamic LENS that integrates information over time, like knowledge tracing models, while simultaneously capturing epistemic uncertainty, like measurement models. An extension of the static LENS model~\cite{christie2023lens}, the model presented here uses a variational autoencoder to represent an inference about student skills as a multidimensional Gaussian distribution in a latent space. It then uses forecasts, combined with Bayesian updating, to integrate observations over time in the manner of a state-space model. While the model does assume a simple Gaussian latent space distribution, the decoder's flexibility enables complex data representations. We show proof-of-concept performance on two datasets. The new model is competitive with other approaches while maintaining an internal representation of epistemic uncertainty that can power adaptive testing and allow educators to understand the confidence associated with the statements the model makes about student knowledge and skills.

\section{Dynamic LENS Model}

The LENS model, introduced in a previous work ~\cite{christie2023lens},  is a VAE-based model that maps student item responses to Gaussian distributions in a latent space, then aggregates observations using Bayes' Rule to produce a posterior belief about a student's latent parameters. Samples are taken from this latent distribution and then `decoded' to make predictions of student behavior for unobserved interactions. 
LENS supports modeling very sparse observations, like those present in personalized practice and computer adaptive tests. 
Whereas in typical VAEs, student responses are batch-encoded and mapped directly to a student-level posterior distribution, here each single observation is separately mapped to a Gaussian in the latent space and is treated as the observation's likelihood. The likelihoods of conditionally independent observations are multiplied together, and then multiplied by the prior (typically, a standard normal with diagonal covariance) to produce a posterior belief. Samples from the posterior are then combined with item embeddings to produce predictions for behavior on new items (see~\cite{christie2023lens} for a complete description of static LENS).
 
In this paper, we develop a dynamic version of LENS to model student knowledge over time: at each timestep, the prior belief is a forecast from the previous timestep. As LENS represents beliefs about students as Gaussian distributions, we adopt the framework of state-space modeling to integrate observations over time using Bayesian updating. A state-space model is a type of latent-variable model in which a belief about a system's state is updated over time, analogous to a Hidden Markov Model but with a continuous latent space. The dynamic LENS model presented here is a recursive state estimator, similar to the Kalman Filter~\cite{pei2019kalmanfilt} but with relaxed assumptions.  

As shown conceptually in Figure~\ref{fig:conceptual}, the model is defined by a state transition model (also called a \textit{forecast}) and an observation model (also called an \textit{encoding}). Let $x_{i,t}$ be the true latent parameters for student $i$ at time $t$, and let $y_{ij,t}$ be student $i$'s behavior in response to item $j$ at time $t$. The system evolves between discrete timesteps according to a transition function $f$:
\begin{eqnarray}
\text{forecast}(x_{i,t-1}) &=& p(x_{i,t} | x_{i,t-1}, u_t)\\
&=& \mathcal{N}(f_m(x_{i,t-1}, u_t), f_v(x_{i,t-1}, u_t)\mathbb{I})
\end{eqnarray}
where $x_{i,t}$ is the state at time $t$ and $u_{t}$ is exogenous control input at time $t$. $f_m$ produces a vector of means, and $f_v$ produces a vector of log-variances. Similarly, a student response to item $j$ is incorporated into an observation function $h$:
\begin{eqnarray}
\text{encoding}(y_{ij,t}) &=& \mathcal{N}(h_m(y_{ij,t}), h_v(y_{ij,t})\mathbb{I})
\end{eqnarray}
The encoding takes the place of the observation equation in state-space modeling, as it defines a probabilistic relationship between the latent state and observations. We combine the forecast and the behavioral encoding by multiplying to produce a posterior belief over the student latent space at time $t$, namely $p(x_{i,t}|x_{i,t-1}, y_{ij,t})$. The behavioral encoding acts as a likelihood function that re-weights the prior in the latent space according to the relative probability of observations given each point in the space.

As is typical of a discrete time state-space model, the dynamic LENS model (Figure~\ref{fig:conceptual}) produces predictions over time via a sequence of time steps $t \in (0, 1, ..., T)$:
\begin{enumerate}
    \item  For a timestep $t > 0$, the past estimate (posterior at $t-1$) is forecast using $f_m$ and $f_v$ to produce the prior distribution at time $t$. The forecast at timestep $t > 0$ is likely to increase uncertainty and shift the mean (due to student knowledge likely changing over time). At time $t=0$, the prior is a standard normal.  
    \item Individual item responses are mapped to parameters of the encoding by $h_m$ and $h_v$. The parameters of the component Gaussians are used to produce the product of multiple encodings in closed-form. These encodings re-weight the student belief space and are treated as likelihoods.
    \item The product of the encodings is multiplied by a pre-existing prior distribution to produce a posterior distribution over student belief space. This posterior will serve as a past estimate for the next time step.
    \item Posterior samples are taken to make predictions at time $t$.
\end{enumerate}

\begin{figure}
    \includegraphics[width=\columnwidth]{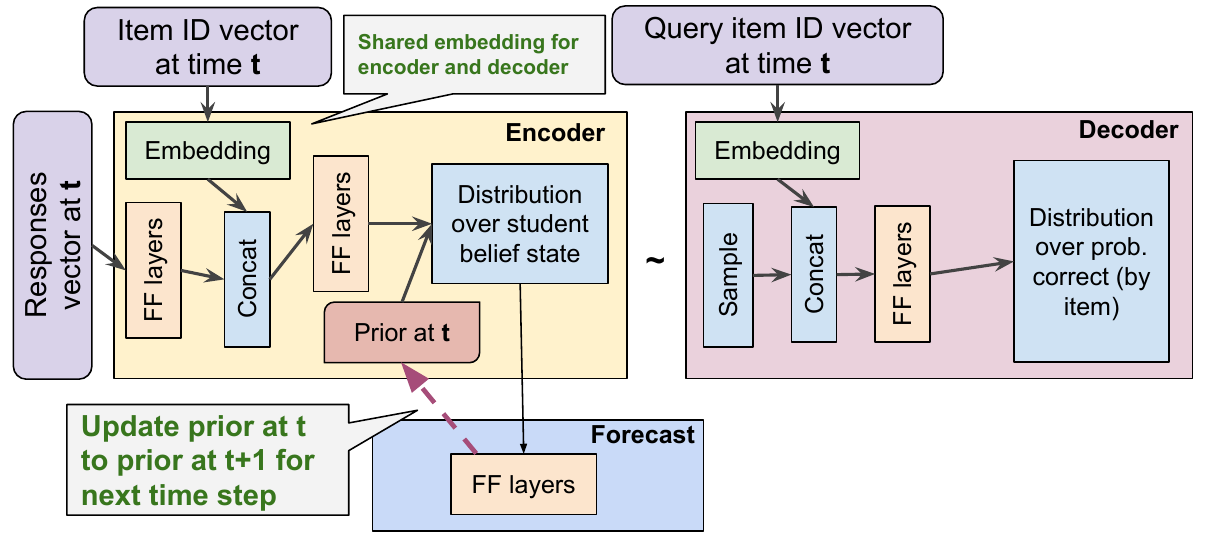}
    \caption{Dynamic LENS architecture. The encoder (left) learns a function for relating student behavior on test items to a distribution over the student latent space. Responses to multiple items are combined together using an analog of Bayesian updating. The decoder (right) samples from the latent space and learns a function to map samples to predictions of future student behavior. When incorporating new observations about the same student at a later time point, the current distribution over the student state is forecasted to form a new prior over the student state.}
    \label{fig:architecture}
    \Description{Architecture diagram for the Dynamic LENS architecture. The figure shows that responses at time $t$ are combined with embedded item ids at time $t$ and then encoded using feed-forward layers. The encoded representations are combined with the prior at time $t$ to give a distribution over the student belief state. The figure shows that embedding of item IDs is shared between the encoder and decoder. The decoder takes item IDs as input, and these embedded items and samples from the student belief state are combined using feed-forward layers to output a distribution over the probability correct by item. A forecast layer is shown that takes input from the distribution over student belief state and uses feed-forward layer to update the prior for the next time step.}
\end{figure}

The functions $\hat{f}_{m}$, $\hat{f}_{v}$, $\hat{h}_{m}$, and $\hat{h}_{v}$ are parameterized neural networks learned from data (see Figure~\ref{fig:architecture} for an architecture diagram). The loss function is the typical VAE loss, which is a weighted sum of the negative log likelihood (NLL) of responses given the posterior and the KL divergence between the posterior and the prior beliefs. The NLL component optimizes for predictive accuracy, while the KL component regularizes the information gain from individual observations. We sum this loss at each timestep.

 \section{Model Performance}

As an initial test of the performance of the Dynamic LENS model, we compared it with four models designed to accumulate information about student skill evolution over time, including two latent variable models and two deep learning models. Our goal was to provide a proof-of-concept evaluation for whether Dynamic LENS was competitive with alternative options and to examine the posterior outputs.
We trained each model, including hyperparameter tuning where appropriate, on both a simulated dynamic CDM dataset and on historical student responses on interim assessments.\footnote{Hyperparameters and a complete architecture diagram for LENS are available \href{https://osf.io/q8pa3/?view_only=a34136a65fef41ef9e1ebf4749ee756a}{in the supplementary materials.}}

\subsection{Comparison Models}

We compared Dynamic LENS with both Elo~\cite{elo, pelanek2016applications} and Bayesian Knowledge Tracing (BKT~\cite{corbett1994knowledge,pardos2013adapting}), latent variable models used to track student progress in intelligent tutoring and formative assessment systems. Elo treats student skills as continuous and updates student skill estimates after each response. Our implementation of the Elo model is based on \cite{elo}, and uses a separate skill parameter for each skill category in the dataset. BKT treats student skills as independent discrete latent variables. We used the \textsc{pyBKT}~\cite{badrinath2021pybkt} implementation with both forgetting and ``multigs'' enabled. 

We also compared Dynamic LENS with standard Deep Knowledge Tracing (DKT) and SAINT, as implemented in the pyKT package~\cite{liupykt2022pykt}. The DKT implementation uses an RNN with LSTM units, similar to the original 2015 paper~\cite{piech2015dkt}. The SAINT model is a transformer-based architecture that considers a context window of previous student responses when predicting the response to the subsequent item~\cite{shin2021saint}.

\begin{figure}
    \centering
    \includegraphics[width=1\linewidth]{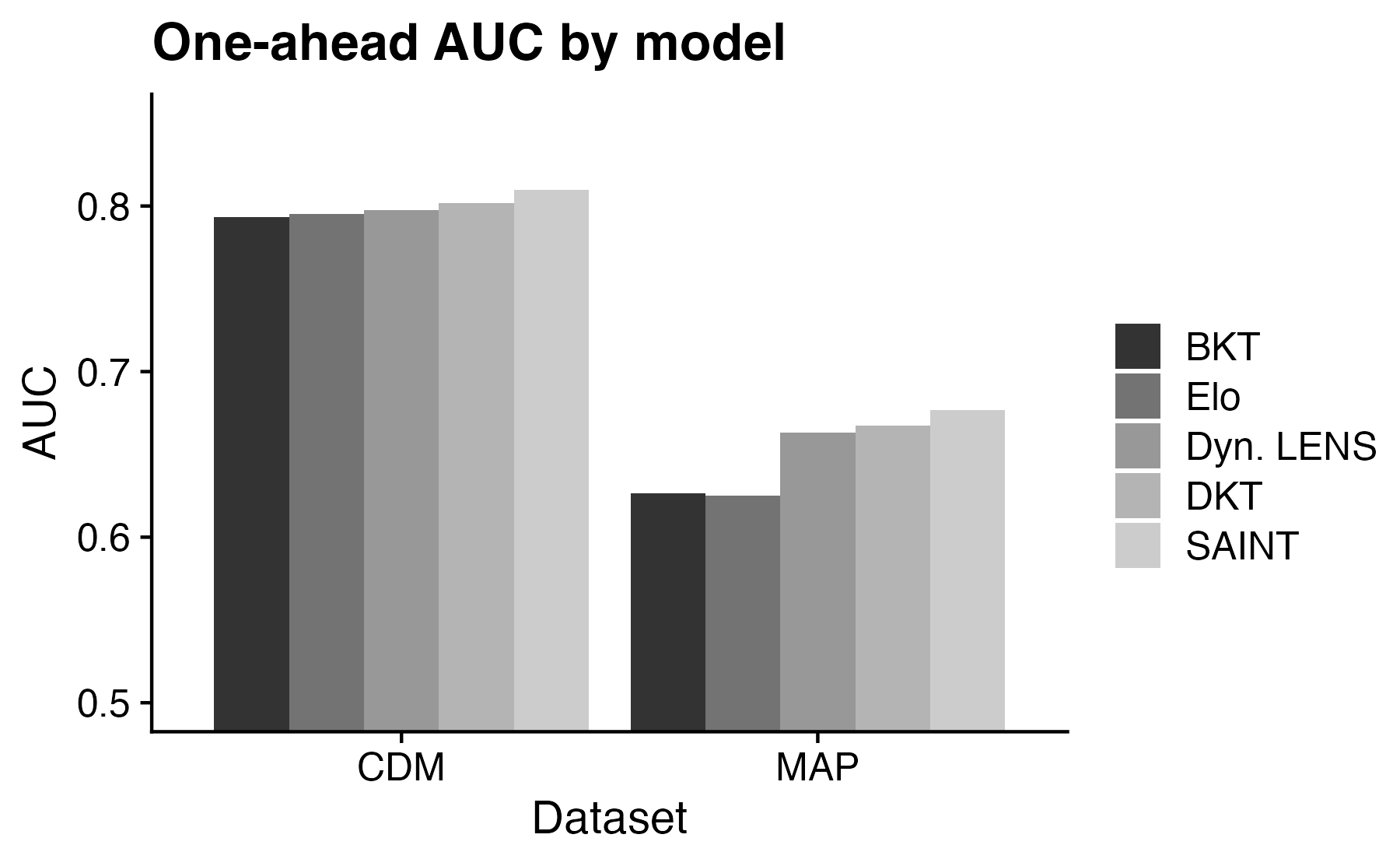}
    \caption{Predictive AUC by model. All models have comparable performance on the CDM dataset. For MAP, lower AUCs are explained by the adaptive nature of the test. LENS performance on MAP falls between that of the simpler models (BKT/Elo) and that of the deep models (DKT/SAINT). }
    \label{fig:auc}
    \Description{Bar chart showing AUC for each model, predicting responses 1-step-ahead for each student.  The figure shows results for both CDM and MAP datasets. AUC for CDM is comparable across all models, with SAINT performing best, then DKT, then LENS, then Elo, and finally BKT. AUC for MAP is lower overall, with BKT and Elo performing slightly worse than LENS and DKT and SAINT performing slightly better than LENS.}
\end{figure}

\begin{figure*}
    \centering
    \includegraphics[width=1\textwidth]{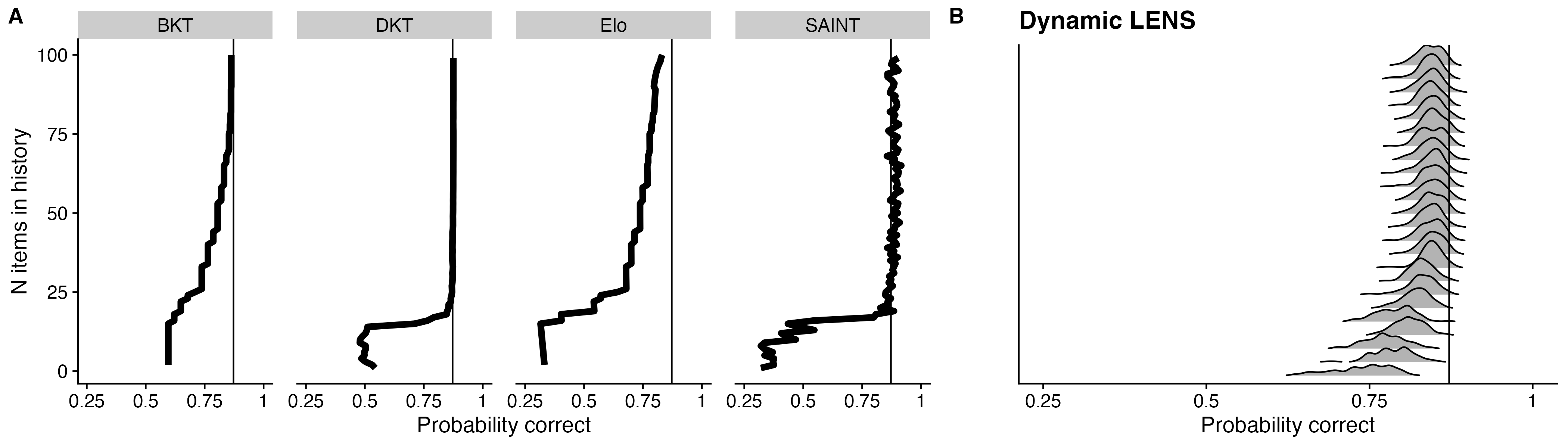}
    \caption{Estimates from each model predicting one simulated student's response to an item given the previous $N$ items. The vertical axis indicates the number of items considered in the student's history. For example, 5 items (bottom row) indicates that only the most recent 5 items are considered, whereas 50 items (middle row) indicates that the 50 most recent items are considered. All models are closer to the true probability correct (gray line) with more information, and the posteriors from Dynamic LENS (B) shows that the model's estimate is more confident with more information. Including information further in the student's history has diminishing returns on both the model's accuracy and level of confidence.}
    \label{fig:single-student-density}
    \Description{The figure shows 5 panels, with each showing the model's prediction that a given student will answer an item correctly. At the bottom, the prediction is made informed by only one historical item response. As the plots progress from bottom to top, more item responses are included in each model's history, until at the top 100 historical items are included. The comparison models show single lines that move towards the true probability, while the Dynamic LENS models show probability distributions that converge to the same point.}
\end{figure*}

\subsection{Datasets}

We examine performance on two datasets that measure student progress over time, one historical and one synthetic. 

\paragraph{MAP Growth Mathematics} The MAP Growth Mathematics assessment is an interim assessment taken by millions of K-12 US students, typically three times a year. The assessment is adaptive, meaning that items are selected sequentially for each student to improve assessment efficiency. We fit each model to data from a cohort of students who completed all six math tests over their 4th and 5th grade years; data were provided in anonymized form and their use is not considered human subjects research by the IRB. Each of the six assessments has roughly $52$ questions.
In total, the dataset contains 157,338 students and approximately $49M$ responses.

\paragraph{Dynamic Cognitive Diagnostic Model} We also constructed a simulated dataset that incorporates student learning into a cognitive diagnostic model (CDM). CDMs treat student skills as binary attributes and typically include conditional dependencies between skills. This dataset contains 10k simulated students based on the skill prerequisite relationships on page 18 of~\cite{mislevy1999bayes}, a well-known analysis of the Mixed Fraction Subtraction dataset from~\cite{tatsuoka1984analysis}. Each simulated student started at time $t=0$ with a configuration of skills, in which the probability of having each skill was $p=0.2$ if the student had all prerequisites and $p=0.05$ if they were missing one or more prerequisites, similar to a DINA model~\cite{delaTorre2009dina}.

Students were then forecast forward in time in discrete time steps. At each time step, students could learn (L) or forget (F) skills independently and with probabilities depending on both the skill value at the previous time step and the presence of prerequisite skills at the current time step; students were more likely to learn skills for which they had the prerequisites. We created a 1000-item item bank, where each item was aligned to 1 of 6 skills; see \href{https://osf.io/q8pa3/?view_only=a34136a65fef41ef9e1ebf4749ee756a}{supplementary materials} for complete dataset generation code. At each of the 100 time steps, students responded to 5 randomly chosen items. In total, the dataset included 5M simulated responses.

\subsection{Results}
We first examine predictive AUC on each dataset when predicting response $N$ for each student in the test set using that student's previous responses $1,\ldots, N-1$. $N$ varies from two to the total number of items in the dataset. As shown in Figure~\ref{fig:auc}, all models have relatively similar AUC for the simulated CDM dataset, with SAINT having a slight edge over all other models. 
For MAP, Dynamic LENS outperforms the simpler BKT and Elo, and is competitive with DKT and SAINT. Overall AUC is much lower for MAP than for the simulated dataset, likely due to the adaptive item selection in the test design.

Unlike the other models in this comparison, Dynamic LENS estimates epistemic uncertainty. Figure~\ref{fig:single-student-density} illustrates the changing estimates of all models when more history is available. The Dynamic LENS posteriors (Fig.~\ref{fig:single-student-density}B) show increasing certainty about the estimate as more information is collected, becoming more concentrated as the number of items in the history increases from one (bottom of plot) to 99.   In a model of a static latent parameter, uncertainty typically decreases as more observations are included in the computation of the posterior distribution.  Here, each row includes both new observations and a forecast between timesteps. The former decreases uncertainty, while the latter typically increases it. Furthermore, each prediction uses information from further back in history. For this reason, uncertainty reaches a stable state after about 40 items and does not decrease further. The plots also demonstrate that individual items can have differential impacts on the estimate, depending on the information each student response provides about the prediction in question.

 \section{Discussion}

In this paper we introduce a model called Dynamic LENS that integrates information across multiple assessments over time. We share proof-of-concept results that this model has competitive predictive performance with existing knowledge tracing models while also providing estimates of epistemic uncertainty. The goal of Dynamic LENS is to integrate information across both formative assessments and other types of interactions, such as homework or standardized tests. Such integration must consider the both the reliability of each piece of new evidence and the confidence of any resulting belief. The architecture of Dynamic LENS makes such integration possible: each response is mapped to a Gaussian distribution, and accumulated to create a posterior belief akin to measurement error, but with greater flexibility. In contrast to measurement models used in educational assessment, Dynamic LENS also avoids imposing strong inductive biases via the structure of the latent space and instead learns a representation from data in a manner that optimizes predictive accuracy.

Unlike most machine learning models, Dynamic LENS represents both aleatoric and epistemic uncertainty in its predictions. Epistemic uncertainty is decreased when more observations are made and is therefore critical for signaling when predictions can be relied upon, or when more data would be likely to change them. This is the machine learning analogue of the notion of reliability in traditional assessment and is not typically captured by Deep Knowledge Tracing models. Quantifying epistemic uncertainty also allows us to perform active learning, as items can be selected according to expected information gain.  Since Dynamic LENS incorporates information from previous assessments, the model could be used to construct new assessments that are personalized for each student to focus on areas with highest uncertainty, similar to current computer adaptive testing systems.

While the primary features of Dynamic LENS are flexibility and maintenance of epistemic uncertainty, predictive power is also critical. Our comparisons showed that the predictive AUC of LENS was very similar to DKT and slightly worse than SAINT. We suspect that performance is due to the frequent probabilistic forecasts required during model fitting. To address this, we are exploring methods to ensure numeric stability over frequent updates with few observations at each time point, or training schedules that gradually expand the history length.

Several considerations must be made before using Dynamic LENS with real students. First, it is important to assess whether Dynamic LENS makes similar predictions for different subgroups of students, as measured by ABROCA~\cite{gardner2019abroca} and MADD~\cite{verger2023madd}, but preliminary (unpublished) analyses look promising.  Second, because of the computational demands of Dynamic LENS versus IRT, using Dynamic LENS for adaptive testing will require investigation into how to approximate information gain efficiently. Finally, there is a need to explore reporting strategies so that the results from the Dynamic LENS model can be shared with all stakeholders - administrators, teachers, parents, and students - in a comprehensible and useful way.

\section{Acknowledgments}
The authors thank NWEA for making anonymized data available for research.

\bibliographystyle{abbrv}

 \end{document}